\documentclass[letterpaper, 10pt, conference]{ieeeconf}      
\pdfoutput=1
\IEEEoverridecommandlockouts 
\usepackage{gensymb} 
\usepackage[utf8]{inputenc}
\usepackage[T1]{fontenc}
\usepackage[english]{babel}
\usepackage{amssymb}
\usepackage{bbm}
\usepackage{graphicx} 
\usepackage{mathtools}
\usepackage{amsmath,amsfonts}
\usepackage{ragged2e}
\usepackage{tikz}
\usepackage{pgfplots} 
\usepackage{pgfgantt}
\usepackage{pdflscape}
\pgfplotsset{compat=newest}
\usepackage{caption}
\usepackage{subcaption}
\usepackage{color}
\usepackage{booktabs,ragged2e}
\usepackage[flushleft]{threeparttable}
\usetikzlibrary{backgrounds}
\usetikzlibrary{calc}
\usepackage{gensymb}
\usepackage{romannum}
\usepackage{xr}
\usepackage{svg}
\usepackage{float}
\usepackage[export]{adjustbox}
\usepackage{array}
\usepackage{amssymb}
\usepackage{cite}

\title{\LARGE \bf
Practical Implementation and Experimental Validation of an Optimal Control based Eco-Driving System}

\author{Vinith Kumar Lakshmanan$^{1}$, Olivier Lemaire$^{2}$, Antonio Sciarretta $^{3}$
\thanks{${1}$, ${3}$ are with the Control, Signal and System Department and ${2}$ is with the Scientific Computing Department, IFP Energies Nouvelles, Rueil Malmaison, France. ${1}$ is also with the ECAV Chair at IFP School.(e-mail: vinith.kumar-lakshmanan@ifpen.fr, antonio.sciarretta@ifpen.fr, olivier.lemaire@ifpen.fr)}}

\begin{document}

\maketitle
\thispagestyle{empty}
\pagestyle{empty}

\begin{abstract}
The main goal of Eco-Driving (ED) is to maximize energy efficiency. This study evaluates the energy gains of an ED system for an electric vehicle, obtained from a predictive optimal controller, in a real-world traffic scenario.  To this end, a Visual driver Advisory System (VAS) in the form of a personal tablet is used to advise the driver to follow a target eco-speed via a screen. Two Renault Zoe electric cars, one equipped with the different modules for ED and one without, are used to perform field tests on a route between Rueil-Malmaison and Bougival in France. Overall, the ED consumed, on average, 4.6~$\%$ less energy than the non-eco-driven car with a maximum of 2~$\%$  change in average speed.
\end{abstract}


\section{Introduction}
Eco-driving aims to minimize the energy consumption of a vehicle. This can be achieved by adjusting the vehicle's speed directly in the case of Autonomous Vehicles (AVs) or indirectly using driver advisory systems. Approaches to ED involve either using heuristic rules such as constant speed and smooth acceleration/deceleration, or rigorous mathematical optimization techniques, realizing the full potential of ED. 

In the past decade, several studies have formulated ED as an Optimal Control Problem (OCP) and solved it using different methods such as Dynamic Programming (DP) \cite{Zeng.2018,Maamria.2018, vandeHoef.2017}, Direct Methods \cite{bae.2019,Vajedi.2016,Dollar.2020}, and Indirect Methods \cite{WANG.2,Wang.2014,Han.CF}. The method used often determines whether the solution is applicable online or offline. While DP gives global optimum solutions, its high computation costs prevent real-time implementation.  Therefore, only information available a priori, such as route characteristics (road slope, speed limit, etc.,) are considered in the OCP. On the other hand, analytical solutions obtained by the indirect method (i.e., Pontryagin's Minimum Principle (PMP)) allow for easy online implementation and subsequently consider real-time traffic information in the OCP.  More generally, the online strategies can involve \textit{a posteriori} analysis of driver behavior to eco-driving strategies, such as the disbanded IFPEN Geco App \cite{DIB2014299} or of a predictive nature involving Model Predictive Control (MPC) framework \cite{MORARI19881}. Such a framework facilitates the derivation of eco-driving strategies based on estimating future external conditions, such as anticipation of traffic and route characteristics, while allowing for modeling assumptions and uncertainties. 

An ED control strategy's performance (energy gains) is typically assessed via simulation testing, Vehicle-In-Loop (VIL) testing, or real-world testing. The predominant research in the ED strategies evaluates their performance in simulation testing, which involves modeling the vehicle and its environment. VIL testing replaces the vehicle model with an actual vehicle while still simulating the interaction environment of the car. Such testing is carried out in a test track with a traffic scenario created through microsimulations. The performance of ED control strategies under different scenarios such as signalized intersections \cite{VIL_intSignal1},  un-signalized intersections \cite{VIL_intSignaless}, on-ramp merging \cite{Mahbub2019ADT}, urban corridors leveraging SPaT information via V2I communications \cite{jihunVIL} have been demonstrated using VIL. 

While the above testing methods allow for cost-effective, safe, and quick performance evaluation, actual performance can only be evaluated through real-world testing. Such testing presents challenges, particularly in requiring computationally fast algorithms that adapt to the dynamic disturbances that surrounding traffic conditions pose. This can be achieved through an analytical framework, which allows fast and explicable solutions, enabling efficient real-time implementation. One such implementation for a specific driving scenario (eco-approach and departure) via advisory systems has been studied in \cite{Mahler2017CellularCO}. However, to the best of the authors' knowledge, experimental testing of ED strategies in real and general traffic conditions has not been tried. 

\begin{figure}[h!]
    \centering
    \includegraphics[scale=.2]{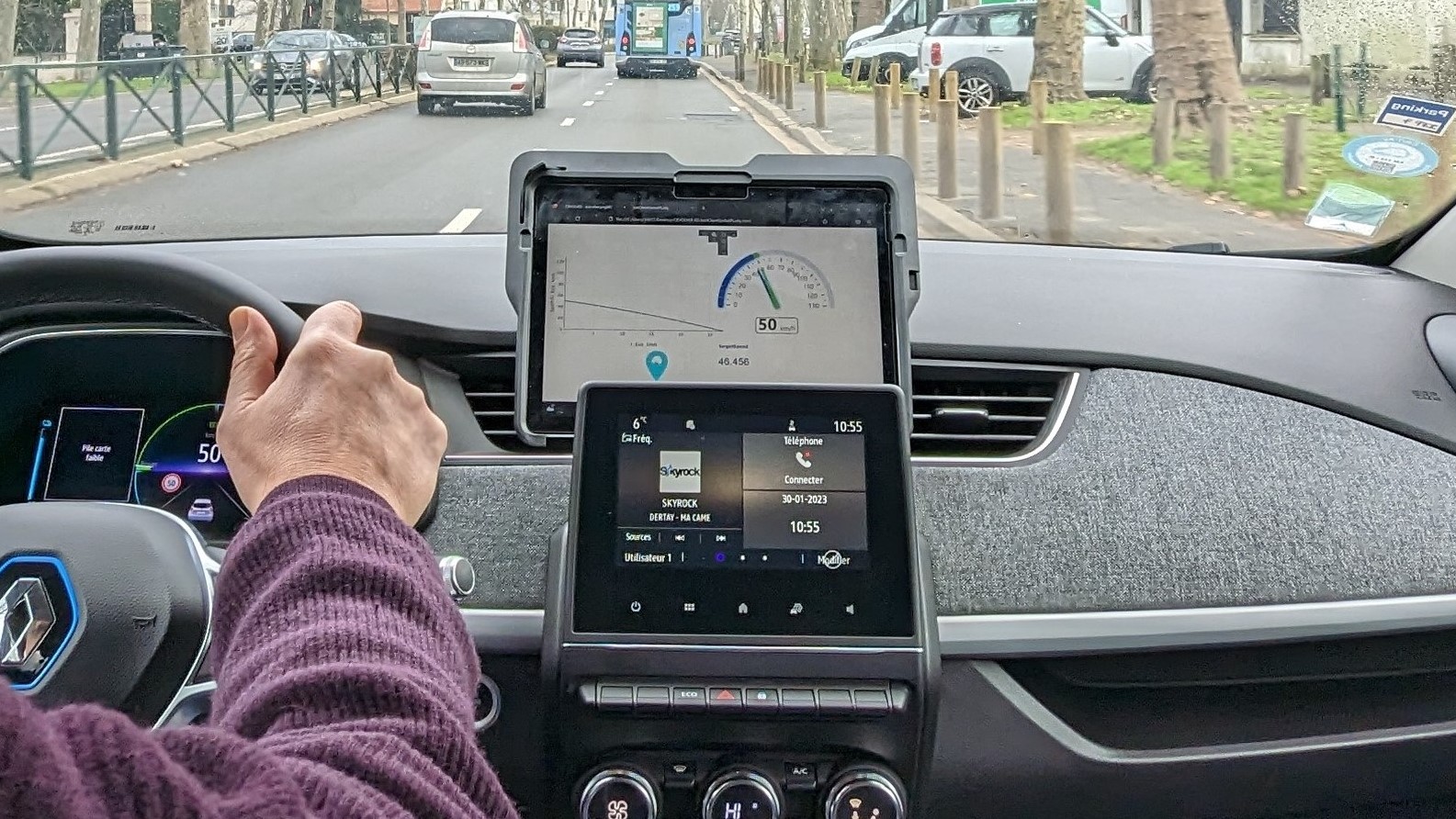}
    \caption{The tablet mounted on the dashboard displays the ED target speed.}
    \label{fig:zoe}
\end{figure}

The main objective of this paper is to evaluate the performance of a predictive ED strategy for an electric vehicle, derived and presented in \cite{Han.CF}, in a real-world traffic scenario. To this end, a Visual driver Advisory System (VAS) in the form of a personal tablet is used to advise the driver to follow a target eco-speed via a screen. This device hosts the ED strategy derived in \cite{Han.CF} and is mounted on the car's dashboard, as shown in Fig.~\ref{fig:zoe}. Analogous to the different modules in an AV, the tablet acts as a motion planning module that takes input from the localization and perception modules. The driver replaces the motion control module. A smartphone and a roof-mounted camera represent the localization and perception modules.  Two Renault Zoe electric cars, one equipped with the various modules for ED and one without, are used to perform field tests on a route between Rueil-Malmaison and Bougival in France. The performance of the ED strategy is evaluated in terms of energy gains against the driver without any eco-driving advice. 

The organization of this paper is as follows: Sect.~\ref{Algorithm} reviews the OCP formulation and solution for an electric vehicle. The following section describes the practical implementation of the ED system and its interaction with the different modules. The experimental procedure and results are presented in Sect.~\ref{results} followed by the conclusion.

\section{Eco-Driving with Pontryagin's Minimum Principle}\label{Algorithm}
This section describes the eco-driving algorithm implemented in the VAS. The algorithm aims to find a velocity profile that minimizes the battery energy consumption of an electric vehicle, going from velocity $v(t)$ to $V$ over a time $T$ and a distance $D$. The initial position is considered zero without loss of generality. 

\subsection{Optimal Control Problem Formulation}

The longitudinal motion of the vehicle is captured by a simple model given by Newton's second law,
\begin{equation}\label{II-vehicle_model}
\begin{aligned}
             \dot{x} &= v, \\
             m \dot{v} &=F_{t}-\left(F_{a}+F_{r}+F_{g}\right)-F_{b}, \\
			ma &= F_{t}-\frac{1}{2}\rho_{a} c_{d} A_{f} v^{2}-m gc_{r} -m g \sin (\alpha(x)) -F_{b}\;,
\end{aligned}
\end{equation}
where $F_t$, $F_a$, $F_r$, $F_g$, and $F_b$ are the traction force given by the powertrain at the wheels, the aerodynamic resistance, rolling resistance, resistance due to gravity and the mechanical braking force, respectively.  The states of the system, $v$ and $x$, represent the velocity and position of the vehicle, respectively. The vehicle's mass is $m$ and $a$ is the acceleration. Considering single-gear transmissions only, the inertial mass is constant and is incorporated into $m$. The parameters that contribute to aerodynamic drag are $\rho_{a}$, $c_{d}$, $A_{f}$, denoting the external air density, aerodynamic drag coefficient, and frontal vehicle area, respectively. The parameters contributing to the rolling and gravity resistances are $c_r$, the rolling resistance coefficient, $g$, the gravitational constant, and $\alpha$, the road slope as a function of the position $x$. 
To facilitate closed-form analytical solutions, the non-linear vehicle model is linearized under the following assumptions:
\begin{itemize}
    \item The resistive forces per unit mass are modeled as a constant $h$. Although a strong assumption, an electric city car generally travels at low speeds, and the error introduced may be limited.
    \item Only regenerative braking is possible  (i.e., $F_b = 0$).
    \item The magnitude of the maximum and minimum acceleration is equal (i.e., $a_{max} = - a_{min}$). 
\end{itemize}
Following the assumptions, the control input is $a = F_t/m - h$ and the linearized vehicle model reads:
\begin{equation}\label{linearvehiclemodel}
\begin{aligned}
            \dot{x} &= v,\\
            \dot{v} &= a = F_t/m - h.
\end{aligned}
\end{equation}

The electric power consumed by the motor from the battery is captured through an approximated closed-form analytical expression for a DC motor \cite{6043133} given by 
\begin{equation}
      P_m = mT_m \omega_m  + p_1 T_m^2,
\end{equation}
where $T_m$ represents the motor torque, $p_1$ represents the battery modelling parameter and $\omega_m$ is the motor speed. Assuming no power link, transmission ($\eta_t =1$), or battery losses, the battery power is obtained as $P_b = P_m$.

Under the above-stated assumption, the OCP for an electric vehicle in the absence of any constraints is formulated as
\begin{equation} \label{OCP}
\begin{aligned}
&\underset{a(\tau)}{\text{min}} \quad J = \int_{t}^{t+T} \left[p_0\left(a(\tau)+h\right)v(\tau) + p_1\left(a(\tau)+h\right)^2\right]\,d\tau,\\
&\text{with state dynamics (\ref{linearvehiclemodel}) and final boundary conditions:} \\  
& \qquad  \mathcal{BC} \triangleq \{x(t+T) = D,v(t+T) = V\},
\end{aligned}
\end{equation}
where $t$ denotes time, $\tau$ denotes the running variable within the optimization horizon, and $p_0 = m/r_w$, where $r_w$ is the radius of the wheel. 
The OCP (\ref{OCP}) is solved using PMP \cite{PMP}, and the detailed solution method is presented in \cite{Han.CF,sciarretta2020energy}. Only the main equation is summarized here. In the absence of any constraints, the optimal trajectory $v^*(\tau)$ is a quadratic function of time and is explicitly obtained as
\begin{equation}\label{II-uncOptV}
\begin{aligned}
    v^*(\tau)=v(t)+\left(-\frac{4 v(t)}{T}-\frac{2 V}{T}+\frac{6 D}{T^2}\right) \tau + \\ \left(\frac{3 v(t)}{T^2}-\frac{6 D}{T^3}+\frac{3 V}{T^2}\right) \tau^2.
\end{aligned}
\end{equation}

A first-state inequality constraint is formulated for the ego vehicle to stay below the maximum speed limit as
\begin{equation}\label{vmax}
  v(\tau) \leq  v_{max}.
\end{equation}
The second constraint is formulated to ensure a constant minimum inter-vehicle distance $s_{min}$ in the presence of a preceding vehicle denoted by $l$. The spacing gap between the ego and the preceding vehicle of length $L$ is defined as
\begin{equation}\label{spacing error dynamics}
\begin{aligned}
      \xi(\tau) &= x_l(\tau) - x(\tau)- L  - s_{min},\\
    \dot{\xi}(\tau) &= v_l(\tau) - v(\tau),
\end{aligned}
\end{equation}
In this scenario, the ego vehicle tries to avoid a rear-end collision with the lead vehicle, imposing state inequality constraints of the type,
\begin{equation}
    \xi(\tau) \geq 0.
\end{equation}
The lead vehicle's motion $x_l(\tau)$ is predicted under a constant acceleration assumption, and the position inequality constraint is rewritten more specifically as 
\begin{equation}\label{cfPosContraint}
  x(\tau)  \leq  x_l(t) + v_l(t)\tau + \frac{1}{2}a_l(t)\tau^2, \quad \tau\in[t,t+T],
\end{equation}
where the lead vehicle's velocity, position, and acceleration are represented by $v_l(t), x_l(t)$ and $a_l(t)$, respectively. Note that $s_{min}$ and $L$ are lumped in $x_l(t)$. 

Only the unconstrained solution described in (\ref{II-uncOptV}) is employed in this study. However, the validity of (\ref{II-uncOptV}) in the presence of a maximum speed constraint (\ref{vmax}) is bounded by 
\begin{equation}\label{f_vmax}
\begin{aligned}
f_1= & \frac{v(t)+V+v_{\max }}{3}  -\frac{D}{T} + \\& \frac{\sqrt{v(t) V+v_{\max }^2-v(t) v_{\max }-V v_{\max }}}{3}  \geq 0,
\end{aligned}
\end{equation}
and in the presence of a lead vehicle constraint (\ref{cfPosContraint}) is bounded by
\begin{multline}\label{f_xi}
    f_{2}=  x_l(t) + v_l(t) \tau + \frac{1}{2}a_l(t)\tau^2 - v(t) \tau -  \left(-\frac{4 v(t)}{T}- \right.\\
\left. \frac{2 V}{T}+\frac{6 D}{T^2}\right) \tau^2 - \left(\frac{3 v(t)}{T^2}-\frac{6 D}{T^3}+\frac{3 V}{T^2}\right) \tau^3 \geq 0.
\end{multline}
The explicit analytical solutions in the presence of the constraints (\ref{vmax}) and/or (\ref{cfPosContraint}) have been solved and presented in \cite{Han.CF,sciarretta2020energy} and the derivation of the conditions (\ref{f_vmax}) and (\ref{f_xi}) are detailed in Appendix~B of \cite{sciarretta2020energy}.

\subsection{Implementation: Model Predictive Control} \label{MPC}
The speed and acceleration of the preceding vehicle generally vary over time. Instead of predicting the lead vehicle's motion (as given by the RHS of (\ref{cfPosContraint})) using its initial measured states and performing a single optimization at the trip's start, the solutions are implemented within a  Model Predictive Control (MPC) framework. Consider a trip of length $D_f$ with duration $T_f$. At each time $t$, based on the current measured system states $x(t)$, $v(t)$, and the relative distance $\xi(t)$ and speed $\dot{\xi}(t)$ to the preceding vehicle, the ego vehicle performs a new optimization over a finite horizon of length $T$ and distance $D$. Only the target speed is advised to the driver. At the next instance of $t$, the states are sampled again. The process is repeated but with updated boundary conditions, $\mathcal{BC} \triangleq \{D = D_f - x(t),  V\}$ and a shorter optimization horizon $ T = T_f - t$.  The terminal speed $V$ is chosen based on traffic lights, stop signs, or changes in speed limits at $D$.  When a green traffic light is detected, or a certain free-flow speed is to be maintained, the final speed $V$ is set to the average traffic speed to avoid traffic disruption. In the case of a red traffic light or stop sign at $D$, the final speed $V = 0$. The lead vehicle's states and acceleration are also subsequently updated as $x_l(t) = \xi(t)$, $v_l(t) =  \dot{\xi}(t)+v(t)$ and the estimated  $a_l(t)$. In the case where the solution (\ref{II-uncOptV}) violates constraint (\ref{vmax}) or (\ref{cfPosContraint}), the final time $T$ is adjusted to satisfy equation (\ref{f_vmax}) or (\ref{f_xi}), respectively. When both constraints are violated, only the position constraint is considered. This fashion of MPC implementation, where the prediction horizon gets smaller as the ego vehicle moves forward, is termed Shrinking Horizon Model Predictive Control. 

\section{Practical Implementation of Eco-Driving}\label{system overview}

This section describes the practical implementation of the ED solution in an electric Renault Zoe ZE-50. The vehicle is equipped with the ED system comprising various localization, perception, and motion planning modules. It is noteworthy that the ED system is independent of the vehicle manufacturer and can easily be set up in any electric vehicle. 

\begin{figure}[h!]
    \centering
    \input{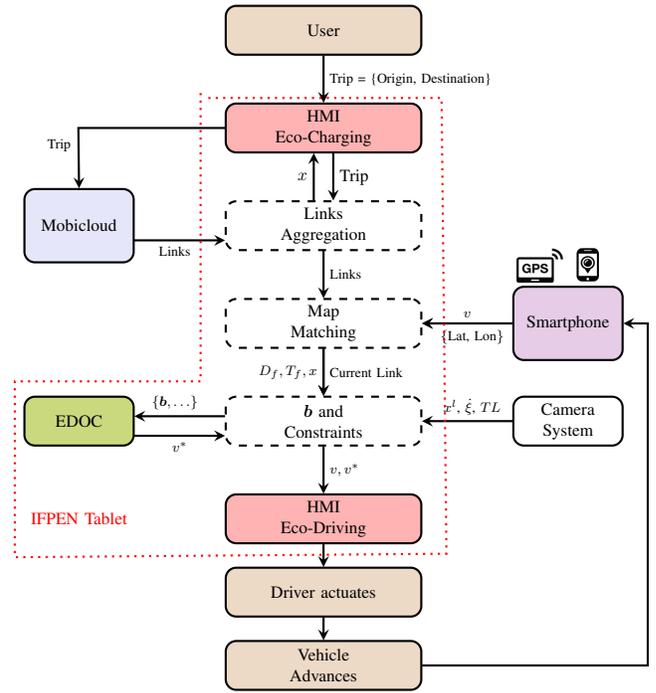}
    \caption{Schematic of the eco-driving advisory system.}
    \label{fig:schematic}
\end{figure}

A smartphone with an in-house developed application obtains the GPS coordinates and the vehicle speed $v(t)$ for localization. The GPS receiver on the smartphone operates at a maximum frequency of 1.15~Hz. A rooftop mountable stereoscopic camera system coupled with an NVIDIA Jetson AGX Xavier controller from the company Visual Behavior is used for perception. Although the ED algorithm can be adapted to pedestrians and other obstacles, the camera system can detect only the presence of a preceding vehicle, measuring its relative position $\xi$ and its relative speed $\dot{\xi}$. The state of a traffic light (i.e., red/green) is also detected. The system has a detection range of approximately 50~m and works at a maximum frequency of 3~Hz. 
A cloud-based computation module, proprietary to IFPEN called the Mobicloud (https://mobicloud.ifpen.com/) server handles the route planning layer. For on-board motion planning, a dedicated personal device tablet is the local computation device that runs the eco-driving algorithm.  The tablet device screen acts as the VAS / HMI, displaying the optimal target speed computed $v^*(3)$ for the driver to follow. This implementation displays the target speed at the third second to compensate for the driver's reaction time. The driver acts as the motion control module and the manufacturer's powertrain control strategy unit is maintained. A conceptual schematic of the various modules and their connections is shown in Fig.~\ref{fig:schematic}. 

Before starting a trip, the driver enters the origin and destination (O/D) in the tablet via the \textit{HMI Eco-Charging}. The O/D pair is communicated to the Mobicloud server that hosts an algorithm named Eco-Charging (EC). This graph-based optimization algorithm computes the energy-optimal route for the given O/D pair composed of several sub-trips or links. The links are obtained based on certain break-point detection criteria, such as a change of speed limit, a new lane, and the presence of an intersection, a traffic light, or a stop sign along the route. Each link is assigned an ID and contains attributes such as its start and end GPS coordinates, speed limit $v_{max}$, link length $D_f$, and predicted final speed $v_f$. The average speed in a link, which is a function of the free-flow and the average traffic speed, computed as a result of the optimization problem in EC, gives the predicted final time $T_f$.

The tablet hosts four functions, namely, \textit{Links Aggregation}, \textit{Map Matching}, \textit{$\mathcal{BC}$ and Constraints}, and  \textit{EDOC}. The links from EC along with its attributes are passed to the \textit{Links Aggregation} function in the tablet. Based on certain heuristic rules, an aggregation process on the links that are too small or where the driver will not perceive any major changes is carried out.

Following the \textit{Links Aggregation}, the driver is displayed the route to follow and can begin the trip. During the trip (i.e., in real-time), the \textit{Smartphone} obtains the $v(t)$ and the GPS coordinate pair of the vehicle. The \textit{Map Matching} function identifies the vehicle's current link using a geometric point-to-curve map matching algorithm \cite{mapmathcing} and its position $x(t)$ along it. The function first involves converting the set of links and the current GPS coordinate pair of the vehicle into the Lambert-Conic coordinate system. Due to limitations in GPS accuracy, the current position is projected onto the set of links along the route to identify the specific link on which it falls—subsequently, the position $x(t)$ along this link is computed via Euclidean distance. 

\begin{figure}
    \centering
    \includegraphics[scale = 0.25]{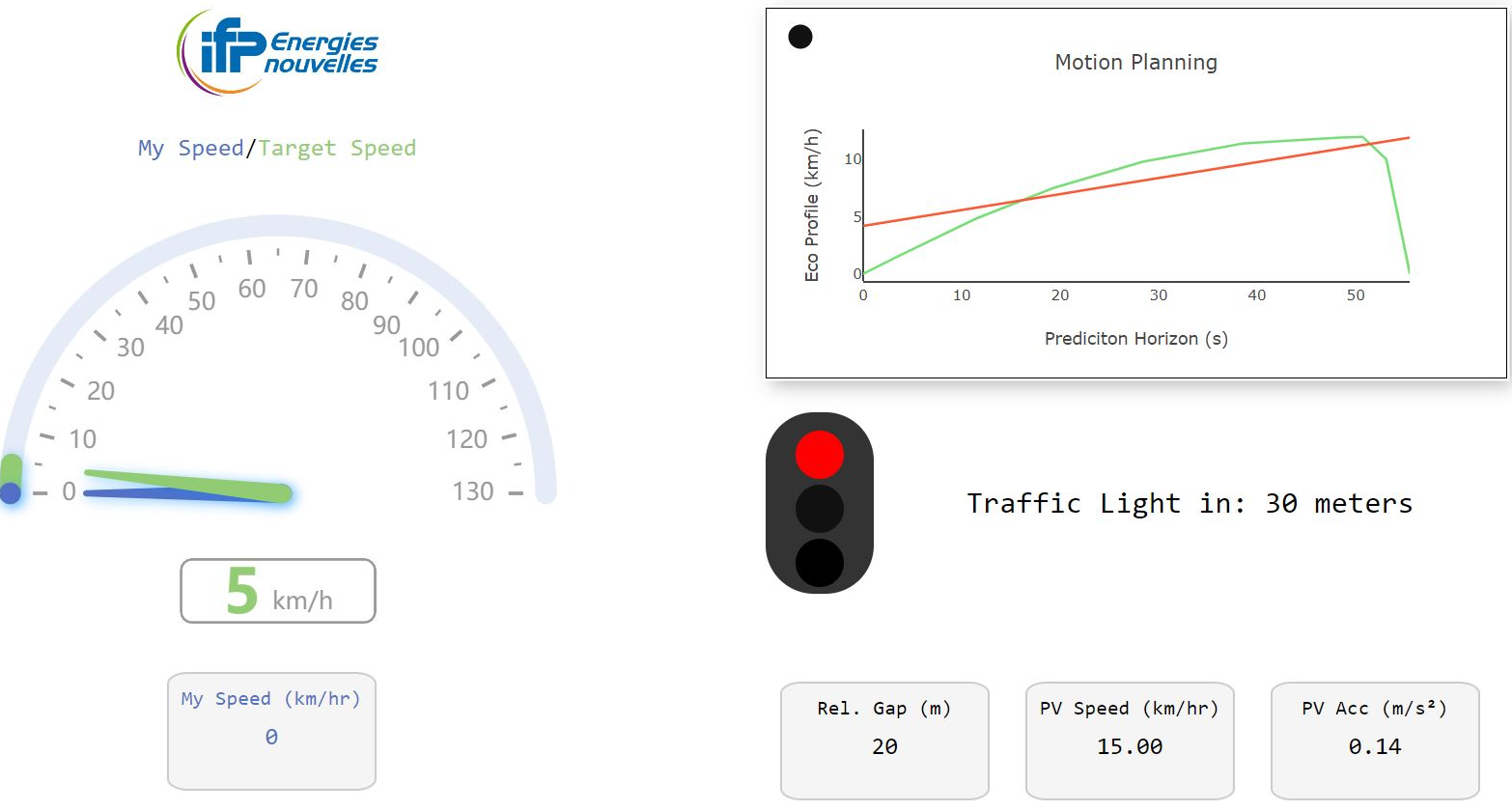}
    \caption{The HMI Eco-Driving.}
    \label{fig:hmied}
\end{figure}

When a traffic light phase is detected, the camera returns a boolean $R$ = 1 (0) for the red light detected (undetected) and $G$ = 1 (0) for the green light detected (undetected). A missed detection of a traffic light phase by the camera system sets $G = 1$ by default. In the presence of a preceding vehicle, the measured relative speed and distance $\xi(t)$ and $\dot{\xi}(t)$ are subsequently transformed to $x_l(t) = \xi(t)$ and $v_l(t) = \dot{\xi}(t) + v(t)$. The acceleration of the preceding vehicle $a_l(t)$ is computed using an exponential weighted moving average using the past six $v_l$ measurements and a weighting factor $\beta = 0.95$. This approach reduces the impact on the acceleration estimation noise, which would otherwise be erratic.  The attributes of the link $D_f, T_f$,  the position $x(t)$ and speed $v(t)$ of the ego vehicle, along with the preceding vehicle and traffic light information, are passed on to the \textit{$\mathcal{BC}$ and Constraints} function. This, in turn, computes the set of $\mathcal{BC}$ and the constraints as detailed in Sect.~\ref{MPC}.  The \textit{EDOC} module computes the ED speed profile for the given $\mathcal{BC}$ and constraints and displays $v^*(3)$ on the \textit{HMI Eco-Driving}, see Fig.~\ref{fig:hmied}. The driver then actuates the acceleration pedal to follow the advised speed and advances the vehicle. Real-time computations, i.e., localization, perception,  map matching, evaluation of  $\mathcal{BC}$ and constraints, and running of the EDOC, are carried out iteratively at a minimum of 1~Hz until the end of the trip.  The communication between the different modules uses a combination of web socket Server-Client (web socket protocol) and TCP server-client (TCP/Socket protocol). 

\section{Experiments and Results}\label{results}
This section describes the experimental procedure and the performance of the eco-driving algorithm in actual traffic conditions. The experiment involved driving two Renault Zoe ZE-50s, one with the ED advisory system described in Sect.~\ref{system overview}, called the Eco-Driven (ED), and the other in stock form called the Human-Driven (HD). The experiment was carried out on a route between Rueil-Malmaison and Bougival in France. The route is approximately 2.3~km with a high traffic light density (total of 9 traffic lights). The two cars start and finish at the exact location, ensuring the same surrounding traffic conditions, such as congestion levels and traffic light status along the route. 

\begin{figure}[h]
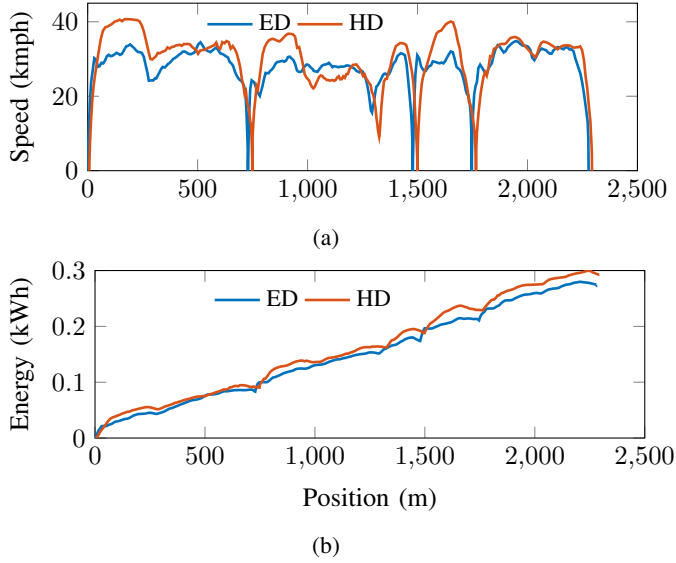

    \begin{subfigure}{\columnwidth}
        \input{figures/Trip8_05012024_speedPosition}
        \caption{}
        \label{fig:speedTrip8}
    \end{subfigure}
    \begin{subfigure}{\columnwidth}
        \input{figures/Trip8_05012024_energyPosition}
        \caption{}
        \label{fig:energyTrip8}
    \end{subfigure}
    \caption{ Speed (a) and energy (b) profiles of the ED and HD for a single trip.}
    \label{fig:trip8}
\end{figure}

Nine trips were carried out on this route to account for the varying traffic conditions. Figure~\ref{fig:speedTrip8} shows the velocity profile of one such trip. In this case, the HD started ahead of the ED and both vehicles started and stopped at a red traffic light. It can be noticed that the HD featured a more aggressive driving behavior, especially during departure from stops, reaching high speeds very early and contributing to increased energy consumption, see Fig.~\ref{fig:energyTrip8}. On the contrary, the ED  tries to follow a target optimal speed computed using the information on the states of the road (speed limit, average traffic speed, and length of the link), signal phase, and preceding vehicle information. This predictive optimal speed allows the ED to anticipate its speed, allowing smoother velocity transitions. Another instance of anticipative behavior by ED, particularly braking,  can be noticed right before the first stop (between 500~m and 1000~m). The ED starts braking earlier than HD, first to reduce its speed to the average traffic speed of the following link (traffic light not detected yet) and then reducing its speed to a complete stop when $R =1$.

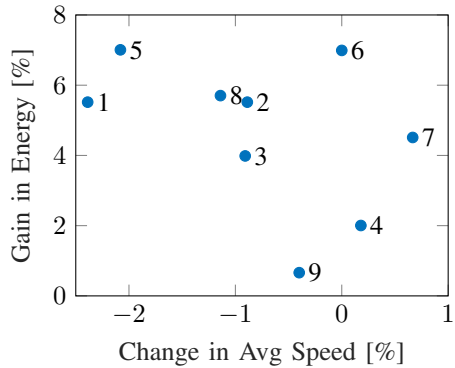
\begin{figure}[h]
    \centering
%
%
\definecolor{mycolor1}{rgb}{0.00000,0.44700,0.74100}%
\begin{tikzpicture}

\begin{axis}[%
width=4.521in,
height=3.566in,
at={(0.758in,0.481in)},
scale = 0.5,
xmin=-2.5,
xmax=1,
xlabel style={font=\color{white!15!black}},
xlabel={Change in Avg Speed [\%]},
ymin=0,
ymax=8,
ylabel style={font=\color{white!15!black}},
ylabel={Gain in Energy [\%]},
axis background/.style={fill=white}
]
\addplot [color=mycolor1, only marks, mark size=2pt, mark=*, mark options={solid, mycolor1}, forget plot]
  table[row sep=crcr]{%
-2.38805970149254	5.51724137931033\\
-0.887772194304857	5.51724137931033\\
-0.90771558245084	3.98550724637681\\
0.17953321364452	2.00429491768074\\
-2.08044382801665	7.00680272108844\\
0	6.99043414275201\\
0.666666666666667	4.51127819548873\\
-1.14009058253943	5.703125\\
-0.400000000000003	0.661157024793384\\
};
\node[right, align=left]
at (axis cs:-2.388,5.517) {1};
\node[right, align=left]
at (axis cs:-0.888,5.517) {2};
\node[right, align=left]
at (axis cs:-0.908,3.986) {3};
\node[right, align=left]
at (axis cs:0.18,2.004) {4};
\node[right, align=left]
at (axis cs:-2.08,7.007) {5};
\node[right, align=left]
at (axis cs:0,6.99) {6};
\node[right, align=left]
at (axis cs:0.667,4.511) {7};
\node[right, align=left]
at (axis cs:-1.14,5.703) {8};
\node[right, align=left]
at (axis cs:-0.4,0.661) {9};
\end{axis}

\end{tikzpicture}%
    \caption{Gain in energy consumption vs change in average speed for the nine trips conducted.}
    \label{fig:energy}
\end{figure}

Figure~\ref{fig:energy} shows the energy gain versus the change in average speed of the ED to the HD.  The energy consumption is calculated \textit{a posteriori} using a backward Renault Zoe vehicle model and the recorded speed profile. The model details are omitted due to space constraints. On average, the ED has an energy gain of 4.6~$\%$ compared to the HD, with a maximum change in average speed of  2~$\%$. The difference and the change in sign of the  average speed occurs due to the different trip times and the starting order between HD and ED, respectively. 
An observation regarding points 4 and 9, which show the lowest energy gain, is their association with high traffic density. The ED and HD vehicles were forced to follow the surrounding traffic patterns, leaving little room for anticipation. 

A post-trip driving style assessment is performed, comparing the actual speed traces with their respective energy-optimal speed. Such an assessment is done in eco-coaching techniques, attributing a score to the driver, enabling them to learn and adapt their driving style to the optimum, see Chap.~8 of \cite{sciarretta2020energy}. The eco-driving score (EDS) is computed as
\begin{equation}
    EDS = \frac{E_D - E_T}{E_T} ,
\end{equation}
where $E_D$ is the energy consumption of the driven profile and $E_T$ is the energy consumption of the optimum profile.
\begin{figure}[h!]
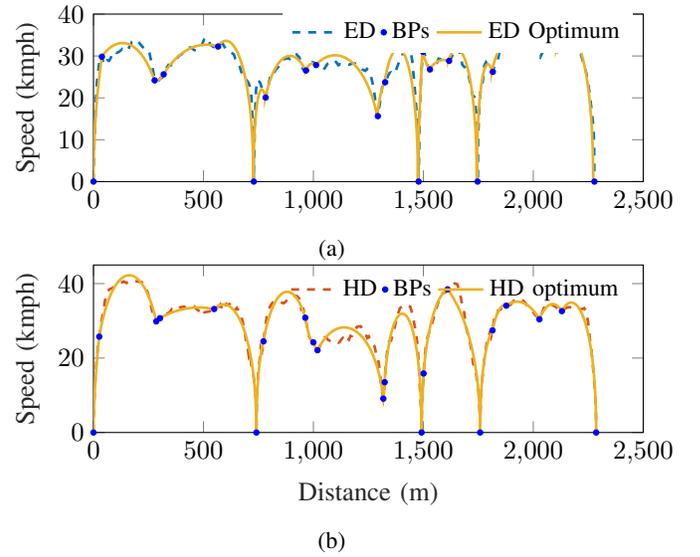

    \begin{subfigure}{\columnwidth}
        \input{figures/trip8_05012024_ecoScoreEd}
        \caption{}
        \label{fig:ecoscoreEd}
    \end{subfigure}
    \begin{subfigure}{\columnwidth}
        \input{figures/trip8_05012024_ecoScoreHd}
        \caption{}
        \label{fig:ecoscoreHd}
    \end{subfigure}
    \caption{Driven and theoretical speed traces for ED and HD. }
    \label{fig:trip8 Eco-coaching}
\end{figure}

\begin{figure}[h!]
    \centering
%
%
\begin{tikzpicture}

\begin{axis}[%
width=4.521in,
height=3.566in,
at={(0.758in,0.481in)},
scale = 0.5,
xmin=0,
xmax=10,
xlabel style={font=\color{white!15!black}},
xlabel={Trip  [-]},
ymin=0,
ymax=0.18,
ylabel style={font=\color{white!15!black}},
ylabel={Score [-]},
axis background/.style={fill=white},
legend style={legend cell align=left, align=left, draw=white}
]
\addplot [color=green, only marks, mark size=2pt, mark=*, mark options={solid, green}]
  table[row sep=crcr]{%
1	0.0861\\
2	0.0773\\
3	0.0705\\
4	0.1155\\
5	0.0562\\
6	0.0392\\
7	0.0644\\
8	0.051\\
9	0.056\\
};
\addlegendentry{ED}

\addplot [color=red, only marks, mark size=2 pt, mark=*, mark options={solid, red}]
  table[row sep=crcr]{%
1	0.1026\\
2	0.1091\\
3	0.0854\\
4	0.0783\\
5	0.1131\\
6	0.0777\\
7	0.0769\\
8	0.0727\\
9	0.0798\\
};
\addlegendentry{HD}

\end{axis}

\end{tikzpicture}%
    \caption{EDS for HD and ED for the nine trips.}
    \label{fig:ecoscore}
\end{figure}
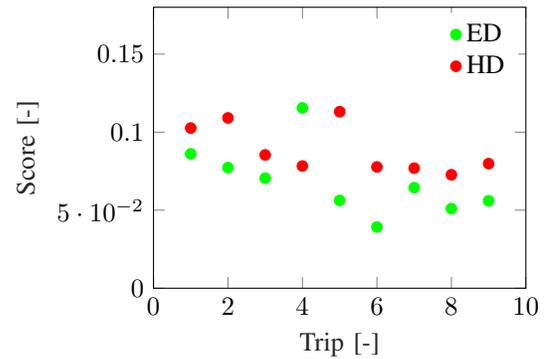
An EDS = 0 indicates a perfect score, and a score close to 1 or more indicates the driver is far from the optimal. The optimal speed profiles are computed \textit{a posteriori} serving to answer the question, \textit{what would have been the optimal speed profile computed at the trip's start if we had access to traffic and signal phase of the entire route?} This is done by analyzing the individual-driven speed profiles and identifying breakpoints due to traffic in addition to those identified at the trip's start by EC. Detecting such breakpoints is done by identifying the speed minima with sufficient prominence; see Chap.~8.2.2 in \cite{sciarretta2020energy} for more details. The energy optimal speed profile is then computed for the new set of breakpoints and compared with the driven ones. Despite efforts to reduce the difference in traffic conditions between the two vehicles, each is subject to different preceding vehicles, traffic queues, etc., along its route. Hence, the driven speed profiles by both HD and ED are individually analyzed.  The breakpoints for one trip are shown in blue in Fig.~\ref{fig:trip8 Eco-coaching}. An example of the driven and optimal speed profiles for one trip is in Fig.~\ref{fig:ecoscoreEd} for ED and Fig.~\ref{fig:ecoscoreHd} for HD. The EDS of all nine trips is shown in Fig.~\ref{fig:ecoscore}. The relatively low scores indicate both drivers tried to act as eco-drivers, while the ED has a lower score than the HD for all but one trip (Trip 4), confirming the benefit of the proposed system. On Trip 4, the HD achieved a lower EDS score due to more favorable traffic conditions allowing it to be closer to its optimum. 


\section{Conclusion}
The eco-driving performance of an electric vehicle ED driving controller, based on optimal control theory, via a driver advisory system was evaluated in real-world traffic conditions. A series of tests with two Renault Zoe cars were conducted on a fixed route, where one car was equipped with the ED advisory system and the one without. The eco-driver followed a target eco-speed computed as a result of a predictive energy optimization.  The performance regarding energy consumption was evaluated using a backward vehicle model and the recorded speed traces. Overall, the ED consumed, on average, 4.6~$\%$ less energy to the HD with a maximum of 2~$\%$ change in average speed. The advised target speed, computed as a result of the predictive optimization, allows the driver to adjust his speed in anticipation of future traffic conditions and reduce energy consumption. This gives the driver in ED a smoother velocity transition than a purely reactive driver in the HD, leading to better energy efficiency.

A  post-trip driving style assessment was also performed, attributing a score (EDS) by comparing the ED and HD speed profiles with their respective theoretical optimum. The relatively low  EDS for ED and HD indicated that both drivers tried to act as eco-drivers; however, the benefit of the proposed system is revealed by the almost systematically lower EDS of the ED with respect to the HD.   

Few limitations of the current implementation provide avenues for future advancements. First, increasing the frequency of the localization module will enable more rapid feedback on the ego vehicle's states. Second, the estimation of the preceding vehicle's acceleration could be refined through the application of more advanced algorithms to mitigate noise. The overall robustness could be enhanced by reducing uncertainties in mapping, such as inaccuracies in the representation of traffic lights or roundabouts in the maps. Future efforts may also involve substituting the driver with an automated control system.

\section*{ACKNOWLEDGMENT}
The authors would like to acknowledge Rémi Agier and Aurélien Cécille from Visual Behavior for their support and the driver participants, namely, Abdelkader Dib, Baptiste Leroux, and Gianluca Nacci.  

\bibliographystyle{IEEEtran}
\bibliography{Lireview.bib}

\end{document}